\documentclass[aps,prb,letterpaper,amsmath,amssymb,reprint]{revtex4-1}
\usepackage{graphicx}

\begin{document}
\title{Modified Becke-Johnson potential investigation of half-metallic Heusler compounds}
\author{Markus Meinert}
\email{meinert@physik.uni-bielefeld.de}
\affiliation{Thin Films and Physics of Nanostructures, Department of Physics, Bielefeld University, D-33501 Bielefeld, Germany}

\date{\today}

\begin{abstract}
We have investigated the electronic structures of various potentially half-metallic Heusler compounds with the Tran-Blaha modified Becke-Johnson (TB-mBJ\-LDA) potential within the density functional theory. The half-metallic gaps are considerably enhanced with respect to values from the Perdew-Burke-Ernzerhof (PBE) functional. In particular the unoccupied densities of states are modified by mBJ\-LDA, and agreement with experiment is considerably worse than for PBE. The agreement of the densities of states can be improved by reducing the Tran-Blaha parameter $c$. However, ground state properties such as the hyperfine fields are more accurately described by PBE than by mBJLDA. Despite its success for ionic and covalent semiconductors and insulators, we conclude that mBJ\-LDA is not a suitable approximation for half-metallic Heusler compounds.
\end{abstract}

\maketitle

\section{Introduction}

The half-metallic Heusler compounds\cite{Heusler, Kuebler83, Galanakis02, Kandpal07} have frequently been considered as ideal electrode materials for spintronic devices. High tunnel and giant magnetoresistance and high spin injection efficiency are expected from half-metals, i.e., materials that have full spin polarization at the Fermi energy. Heusler compounds are ternary  intermetallic compounds with the general chemical formula $X_2YZ$, where $X$ and $Y$ are transition metals, and $Z$ is a main group element. They form the cubic L2$_1$ structure, which has inversion symmetry and belongs to space group Fm$\bar{3}$m. Some of the closely related inverse Heusler compounds with the Hg$_2$CuTi prototype structure (space group F$\bar{4}$3m) have only recently been discovered to exhibit half-metallic ferromagnetism as well.\cite{Luo07, Liu08} The half-metals from both classes of compounds follow the Slater-Pauling rule, which relates the magnetic moment $m$ (given in $\mu_\mathrm{B}$ per formula unit (f.u.)) and the number of valence electrons $N_\mathrm{V}$ via $m = N_\mathrm{V} - 24$.\cite{Galanakis02}

Most electronic structure studies of Heusler compounds are based on the Kohn-Sham framework of density functional theory\cite{HK, KS} (DFT), which is today the main tool to obtain the electronic structure of solids. However, a well-known failure of this framework is the underestimation of band gaps. This is closely related to a missing derivative discontinuity $\Delta_\mathrm{xc}$ in the approximate exchange-correlation (xc) functionals. However, the Kohn-Sham gap $\epsilon_\mathrm{g}$ differs from the true gap $E_\mathrm{g}$ by this discontinuity even for the exact xc functional, which has to be computed and added to the Kohn-Sham gap "by hand".\cite{Gruening06} This is in principle also true for the gap of half-metals.\cite{Eschrig01}

The appropriate framework to discuss band gaps is the many-body perturbation theory, e.g., within the $GW$ approximation.\cite{Aryasetiawan98} Unfortunately, this approach is computationally very expensive. Tran and Blaha recently proposed an alternative, equally accurate and computationally cheaper method to obtain the gap directly as differences of Kohn-Sham eigenvalues: they modified the Becke-Johnson exchange potential\cite{BJ06} with an additional parameter $c$, so that it reads\cite{TB09}
\begin{equation}
v_{x,\sigma}^\mathrm{mBJ}(\mathbf{r}) = c v_{x,\sigma}^\mathrm{BR}(\mathbf{r}) + (3c -2) \frac{1}{\pi} \sqrt{\frac{5}{12}} \sqrt{\frac{2 t_\sigma(\mathbf{r})}{n_\sigma(\mathbf{r})}},
\end{equation}
where $n_\sigma(\mathbf{r})$ is the spin-dependent electron density and $t_\sigma(\mathbf{r})$ is the spin-dependent kinetic-energy density. $v_{x,\sigma}^\mathrm{BR}(\mathbf{r})$ is the Becke-Roussel potential, which models the Coulomb potential created by the exchange hole.\cite{BR89} Due to the kinetic-energy dependent term in the mBJ potential, it reproduces the step-structure and derivative discontinuity of the effective exact exchange potential of free atoms.\cite{Armiento08} The parameter $c$ was proposed to be determined self-consistently from the density by
\begin{equation}\label{eq:c}
c = \alpha + \beta \, \left( \frac{1}{V_\mathrm{cell}} \int_\mathrm{cell} \frac{| \nabla n(\mathbf{r}') |}{n(\mathbf{r}')}  \mathrm{d}^3 r' \right)^{1/2},
\end{equation}
with two parameters $\alpha, \beta$, which have been chosen to fit the band gaps of a broad range of solids. It can be related to the dielectric response of the system.\cite{Marques11, Krukau08} $c$ increases with the gap size and has a typical range of 1.1--1.7.\cite{TB09} The mBJ potential has been proposed to be combined with LDA correlation (mBJ\-LDA). Its particular merits and limits have been reviewed by Koller \textit{et al.}\cite{Koller11}  In a recent paper, Koller \textit{et al.} have suggested a new and more balanced parametrization of $c$, based on a larger test set of solids.\cite{Koller12} This reparametrization gives, however, rather similar results as the original parametrization for small gap materials. Making use of the kinetic-energy density, mBJ is formally a meta-GGA potential.\cite{Mattson02}

In a recent paper, Guo and Liu have used the mBJ\-LDA potential to investigate the half-metallic ferromagnetism of zinc blende transition metal pnictides and chalcogenides. They found that mBJ\-LDA enhances the half-metallic gaps significantly with respect to conventional DFT calculations.\cite{Guo11} In the present paper, we aim to investigate if the half-metallic gap of Heusler compounds is enhanced with mBJ\-LDA, and if such an enhancement leads to an improved description of the electronic structure.

\section{Computational details}

All calculations in this work are based on the full-potential linearized augmented-plane-wave (FLAPW) method. The mBJ\-LDA calculations are done with the \textsc{elk} code.\cite{elk} The mBJ exchange potential is available through an interface to the \textsc{Libxc} library.\cite{Libxc} $\Gamma$-centered $21 \times 21 \times 21$ $\mathbf{k}$-point meshes are used with 286 points in the irreducible wedge for Heusler compounds and 506 points for the inverse Heusler compounds. A gaussian smearing of 1\,mHa is applied in all calculations. The muffin-tin radii are 2.0\,bohr, and the momentum cutoff for the plane-wave expansion is $k_\mathrm{max} = 4.0$\,bohr$^{-1}$. The angular momentum expansion of potential and wavefunctions is taken to $l_\mathrm{max} = 10$. The mBJ exchange potential is coupled with the Perdew-Wang LDA correlation.\cite{PW92} For comparison with results from a generalized gradient approximation, we choose the PBE functional,\cite{PBE} which is the standard functional in most current studies of Heusler compounds. All calculations are based on experimental lattice constants.

\section{Results}

\begin{table}[t]
\caption{\label{tab:results}Magnetic moments and (half-metallic) band gaps computed with PBE and mBJ\-LDA (marked as mBJ) at the given experimental lattice constants.\footnote{Lattice constants and experimental magnetic moments are taken from References \onlinecite{Kandpal07, Balke06, Kuebler83, Jiang01, Kumar08, Endo76, Slebarski00, Qian11, Liu08, Luo07}.} Moments are given in $\mu_\mathrm{B}$\,/\,f.u., gaps are given in eV, lattice constants are given in \AA{}. The parameter $c$ is dimensionless. Asterisks mark gaps, which are above or below the Fermi energy.}
\begin{ruledtabular}
\begin{tabular}{ l l l l l l l l l }
 & $N_\mathrm{V}$ & $a_\mathrm{exp}$	& $m_\mathrm{exp}$ & $m^\mathrm{PBE}$ & $E_\mathrm{g}^\mathrm{PBE}$ & $m^\mathrm{mBJ}$ & $E_\mathrm{g}^\mathrm{mBJ}$ & $c$ \\ \hline
\multicolumn{9}{c}{Co$_2$YZ Heusler compounds}  \\ 
Co$_2$TiAl & 25 & 5.85 & 0.74 & 1.00 & 0.40 & 1.00 & 1.11 & 1.12 \\ 
Co$_2$TiSn & 26 & 6.08 & 1.96 & 2.00 & 0.47 & 2.00 & 1.16 & 1.17 \\ 
Co$_2$VAl & 26 & 5.72 & 1.95 & 2.00 & 0.36* & 2.00 & 0.65 & 1.13 \\ 
Co$_2$ZrSn & 26 & 6.25 & 1.81 & 2.00 & 0.50 & 2.00 & 1.50 & 1.16 \\ 
Co$_2$CrGa & 27 & 5.81 & 3.01 & 3.04 & 0.39* & 3.00 & 1.06 & 1.18 \\ 
Co$_2$MnAl & 28 & 5.75 & 4.04 & 4.03 & 0.61* & 4.04 & 1.29* & 1.14 \\ 
Co$_2$MnSi & 29 & 5.65 & 4.97 & 5.00 & 0.81 & 5.00 & 1.42 & 1.15 \\ 
Co$_2$MnGe & 29 & 5.75 & 4.93 & 5.00 & 0.57 & 5.00 & 1.49 & 1.19 \\ 
Co$_2$MnSn & 29 & 5.98 & 5.08 & 5.03 & 0.39* & 5.04 & 1.36* & 1.19 \\ 
Co$_2$FeAl & 29 & 5.73 & 4.96 & 4.99 & 0.06* & 5.00 & 0.75 & 1.14 \\ 
Co$_2$FeGa & 29 & 5.74 & 5.04 & 5.02 & 0.02* & 5.00 & 0.80 & 1.20 \\ 
Co$_2$FeSi & 30 & 5.64 & 5.97 & 5.47 & 0.11* & 5.79 & 0.82* & 1.16 \\ 
Co$_2$FeGe & 30 & 5.74 & 5.90 & 5.63 & 0.09* & 5.98 & 0.90* & 1.20 \\ \hline

\multicolumn{9}{c}{other Heusler compounds}  \\ 
Mn$_2$VAl & 22 & 5.92 & 1.94 & 2.00 & 0.28 & 2.00 & 0.48 & 1.09 \\ 
Mn$_2$VGa & 22 & 5.91 & 1.88 & 1.99 & 0.02* & 2.00 & 0.27 & 1.15 \\ 
Fe$_2$VAl & 24 & 5.76 & 0.00 & --- & --- & --- & 0.31 & 1.12 \\ 
Fe$_2$VGa & 24 & 5.78 & 0.00 & --- & --- & --- & 0.39 & 1.17 \\ 
Fe$_2$TiSn & 24 & 6.09 & 0.00 & --- & --- & --- & 0.69 & 1.16 \\ 
Ru$_2$MnSb & 28 & 6.20 & 4.40 & 4.03 & 0.28* & 4.06 & 0.44* & 1.19 \\ 
Ni$_2$MnSn & 31 & 6.05 & 4.05 & 4.03 & --- & 4.17 & --- & 1.19 \\ 
Cu$_2$MnAl & 32 & 5.95 & 3.60 & 3.51 & --- & 3.50 & --- & 1.13 \\ 
Cu$_2$MnSn & 33 & 6.17 & 4.11 & 3.86 & --- & 3.91 & --- & 1.19 \\ 
\hline

\multicolumn{9}{c}{inverse Heusler compounds}  \\ 
Cr$_2$CoGa & 24 & 5.80 & 0.35 & 0.08 & 0.19* & 0.03 & 0.66* & 1.17 \\
Mn$_2$CoAl & 26 & 5.84 & 1.95 & 2.00 & 0.43 & 2.00 & 0.68 & 1.12 \\ 
Mn$_2$CoGe & 27 & 5.80 & 2.99 & 3.00 & 0.36 & 3.00 & 0.76 & 1.18 \\ 
Fe$_2$CoSi & 29 & 5.65 & 4.99 & 4.96 & --- & 5.00 & 0.57 & 1.16 \\ 
\end{tabular}
\end{ruledtabular}
\end{table}

\subsection{Gaps and magnetic moments}

\begin{figure*}[t]
\includegraphics[width=\textwidth]{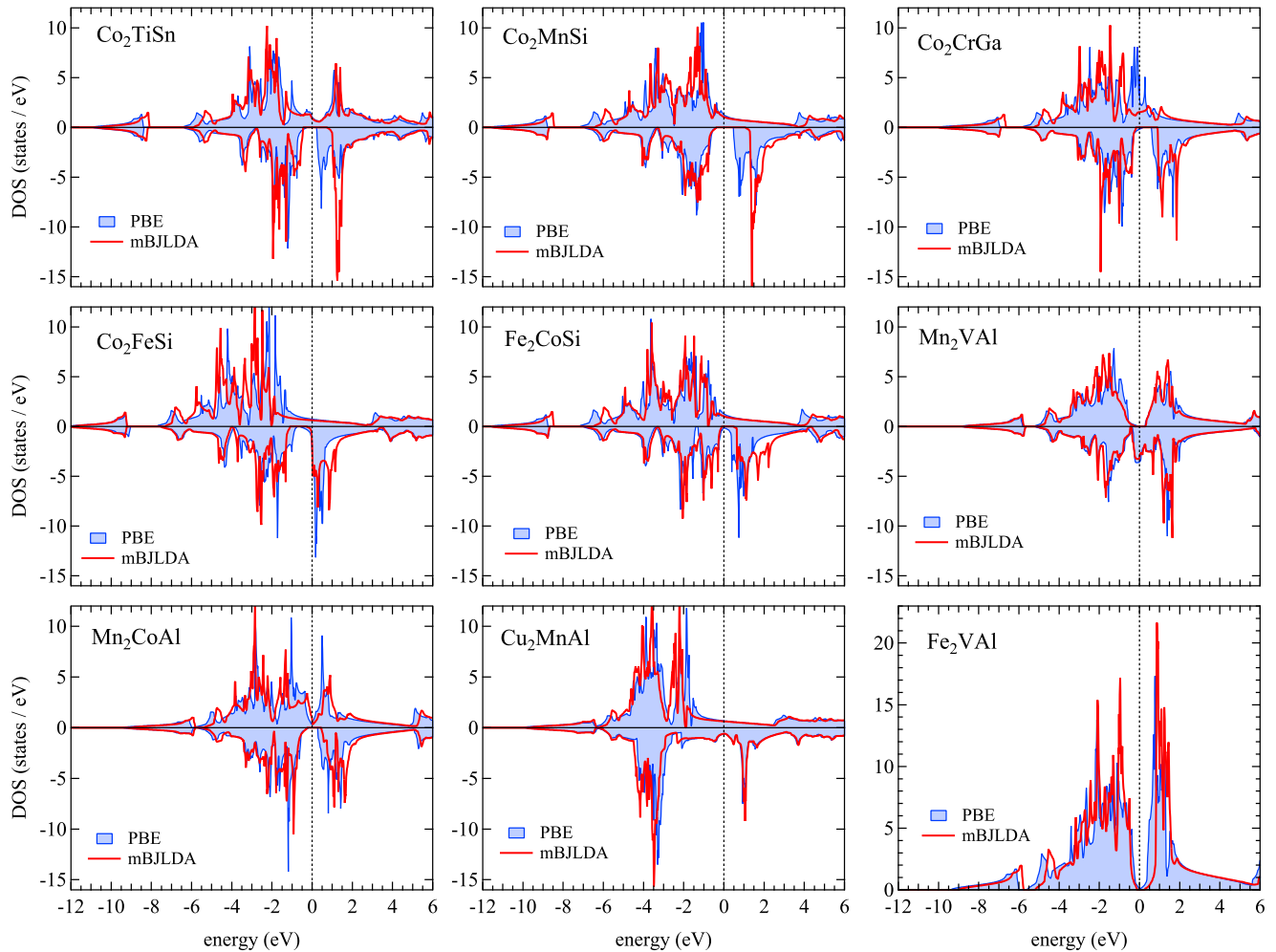}
\caption{\label{fig:DOS}Density of states plots for selected Heusler and inverse Heusler compounds. Shaded blue areas correspond to PBE calculations, solid red lines represent mBJ\-LDA calculations. Spin majority states are positive, minority states are negative. The energy scale is given with respect to the Fermi energy.}
\end{figure*}

We have chosen the Heusler compounds for our study along the lines of Ref. \onlinecite{Kandpal07}, and we added some (inverse) Heusler compounds of current interest.\cite{Qian11, Luo07, Liu08} The main results of our calculations are collected in Table \ref{tab:results}, which displays the magnetic moments and band gaps within PBE and mBJ\-LDA, and the Tran-Blaha parameters $c$. Notably, the $c$ parameter is always in the range 1.09--1.20. The corresponding density of states (DOS) plots are shown in Fig. \ref{fig:DOS}.

We observe that all materials that are half-metallic with PBE are half-metallic in mBJ\-LDA as well, with an increased gap. Some cases which have a gap in PBE with the Fermi energy located outside the gap (marked by asterisks), become half-metals in mBJ\-LDA (Co$_2$VAl, Co$_2$CrGa, Co$_2$FeAl, Co$_2$FeGa, Mn$_2$VGa). Co$_2$MnAl, Co$_2$MnSn, Co$_2$FeSi, Co$_2$FeGe and Ru$_2$MnSb have a larger gap in mBJ\-LDA, but the Fermi energy remains outside the gap.

Mn$_2$VAl and Mn$_2$VGa are ferrimagnetic Heusler compounds. Mn$_2$VAl is characterized as a half-metal with a majority gap by PBE, and mBJ\-LDA has only weak influence on the size of this gap. This goes along with the smallest value of $c$ among the materials studied here. Mn$_2$VGa has a pseudogap around the Fermi level with PBE. A gap is opened by mBJ\-LDA, and the Fermi energy is located within the gap. Fe$_2$VAl, Fe$_2$VGa, and Fe$_2$TiSn, three semimetals or zero gap semiconductors within PBE, are predicted to be semiconductors by mBJ\-LDA.

Ni$_2$MnSn, Cu$_2$MnAl, and Cu$_2$MnSn were included in this study to observe the influence of mBJ\-LDA on the magnetic moment and the exchange splitting of ferromagnetic Heusler compounds that do not have a gap at all. In all three cases the magnetic moments of PBE and mBJ\-LDA are close; while the agreement with experiment is very good for Ni$_2$MnSn, both approximations underestimate the moment of Cu$_2$MnAl and Cu$_2$MnSn. This is surprising, seeing that mBJ\-LDA predicts too large moments for Fe (2.49\,$\mu_\mathrm{B}$) and Ni (0.74\,$\mu_\mathrm{B}$) and too large exchange splittings at the same time.\cite{Koller11} Notably, PBE (and LDA) already predict too large exchange splittings, although the moments are accurate.\cite{Schilfgaarde06} The small influence of mBJ\-LDA on the magnetic moment in Cu$_2$MnAl and Cu$_2$MnSn may be associated with the localized character of the magnetic moment.\cite{Kuebler83} Thus, an increase of the exchange splitting does not lead to an enhanced magnetic moment. In contrast, Fe and Ni have more itinerant character, so the moments depend strongly on the magnitude of the exchange splitting.

The inverse Heusler compound Cr$_2$CoGa has been predicted as a nearly fully compensated ferrimagnet.\cite{Galanakis11} It has a small gap slightly above the Fermi energy. In the mBJLDA description, this gap is further enlarged and the Fermi energy moves closer to the gap, thereby further reducing the magnetic moment towards zero. Experimentally, a non-zero moment is observed,\cite{Qian11} which may be due to atomic disorder. Mn$_2$CoAl is a ferrimagnet, and it is predicted to be a spin gapless semiconductor\cite{Wang08} by PBE. This prediction has recently been confirmed experimentally.\cite{Ouardi12} This is also the case within mBJ\-LDA, and the minority spin gap is only slightly enhanced (similar to the case of Mn$_2$VAl). Mn$_2$CoGe is a half-metal in both approximations, but mBJ\-LDA considerably enhances the minority gap. Fe$_2$CoSi is described by PBE as a conventional ferromagnet with a pseudogap in the minority spin channel around the Fermi energy. mBJ\-LDA opens a sizeable gap and predicts Fe$_2$CoSi to be a half-metal. 

We note that the magnetic moments inside the muffin-tin spheres are increased with mBJ\-LDA in all cases, which is compensated by an antiparallel interstitial moment (for ferromagnets) or by the antiparallel alignment of the muffin-tin moments (for ferrimagnets). The integer magnetic moments are protected by the half-metallic gaps.

\subsection{Densities of states}

In Figure \ref{fig:DOS} we compare the DOS from PBE and mBJ\-LDA calculations for some selected compounds. We observe that the effect of mBJ\-LDA (compared to PBE) is very material-dependent and non-trivial. In all cases we observe that the energy range of the occupied \textit{d} states is compressed; the \textit{d} band minima are raised and the states close to the Fermi energy are lowered. The exchange-splittings of the occupied \textit{d} states are enhanced in all cases. The low-lying \textit{s} states from the \textit{sp}-element are shifted up in some cases, or remain at the PBE position. We further note that the effects on spin majority and minority states are quite different. The occupied minority states are least affected in most cases, while the occupied majority states and the unoccupied minority states show somewhat larger changes.

The enhancement of the half-metallic gaps is visible for all compounds. Mn$_2$VAl shows only little changes with mBJ\-LDA compared to PBE, which is partly due to the low value of $c$. In the case of Co$_2$FeSi we see that the Fermi energy is located slightly above the bottom of the minority conduction band minimum in both cases.

\subsection{Dependence on $c$}
\begin{figure}[t]
\includegraphics[width=8.9cm]{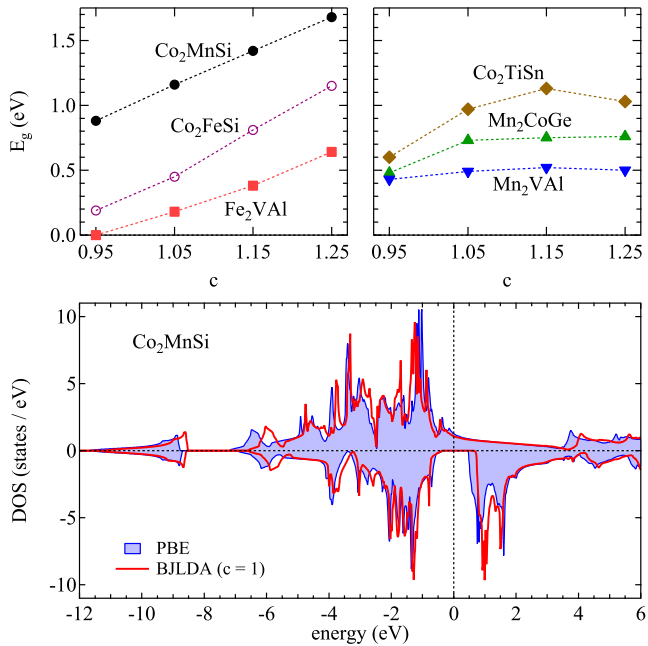}
\caption{\label{fig:c_dependence}Top: Dependence of the half-metallic and semiconducting gaps on the mBJ model parameter $c$. Co$_2$FeSi (open symbols) is not a half-metal for the plotted range of $c$. Bottom: DOS plots for Co$_2$MnSi with the PBE and BJLDA ($c=1$) approximations.}
\end{figure}
To estimate the influence of the mBJ model parameter $c$, we calculated the gaps of Co$_2$TiSn, Co$_2$MnSi, Co$_2$FeSi, Mn$_2$VAl, Fe$_2$VAl, and Mn$_2$CoGe with different fixed values of $c$. The results are displayed in Figure \ref{fig:c_dependence} (top). We can identify two classes of dependencies on $c$ within the range of our investigation: an approximately linear one (Fig. \ref{fig:c_dependence} top left), and one that saturates at rather low $c$ values (Fig. \ref{fig:c_dependence} top right). Remarkably, the three compounds belonging to the second class are ferrimagnets, whereas the compounds in the first class are ferromagnets or nonmagnetic (Fe$_2$VAl). In contrast to semiconductor and insulator gaps,\cite{TB09} the half-metallic gap does not necessarily grow monotonously as $c$ is increased. 

Setting $c\approx 0.95$ nearly restores the PBE values for the gaps in most cases. Also the densities of states agree very well with those from PBE with $c=0.95$. This agrees with the finding by Koller \textit{et al.}, that the magnetic moment of Fe can be tuned to the PBE value with $c$ slightly lower than unity.\cite{Koller11} However, they also mention that the calculated hyperfine fields at the Fe sites are much worse with a small $c$, which indicates subtle differences in the description of the electronic structures, in particular of the \textit{s} states. The original Becke-Johnson exchange potential (i.e., $c = 1$) gives band structures very similar to those obtained with PBE, but with a slightly larger gap. A corresponding DOS plot is shown exemplarily for Co$_2$MnSi in Fig. \ref{fig:c_dependence} (bottom). This agrees with earlier calculations with the BJLDA potential of semiconductor and insulator gaps.\cite{Tran07}

It is well-known that the local xc functionals (both LDA and GGA) fail to predict the ground state magnetic moment of Co$_2$FeSi correctly.\cite{Balke06} mBJ\-LDA does not resolve this problem with the $c$ value from Eq. \ref{eq:c}. With $c=1.35$ a just half-metallic ground-state can be obtained with the Fermi energy at the minority conduction band minimum, and the magnetic moment is $6\,\mu_\mathrm{B}$, in accordance with experiments.\cite{Balke06} As we discuss in the following, such a high value of $c$ is rather unrealistic.

\subsection{Comparison to experiments}
The experimental determination of the half-metallic gap is notoriously difficult. Only indirect evidence for the presence and the size of such gaps is available. This evidence comes from tunnel spectroscopy of magnetic tunnel junctions with amorphous barriers\cite{Kubota09, Sakuraba10,Oogane09} or from x-ray absorption spectroscopy.\cite{Kallmayer09, Klaer09} For Co$_2$MnSi, a combination of the minority spin flip gap from tunnel spectroscopy of $0.25 - 0.35$\,eV and the position of the x-ray absorption maximum of Co at its L$_3$ edge of $0.9\pm0.1$\,eV above $E_\mathrm{F}$ gives an upper boundary of $1.2\pm0.1$\,eV for the minority gap. Note, however, that the Co \textit{d}DOS maximum is about 0.3\,eV above the minority conduction band minimum in the PBE and mBJ\-LDA calculations; the gap is therefore smaller than $0.9\pm0.1$\,eV. Thus, considering the available experimental data, the mBJ\-LDA value for the minority gap is considerably too large and the PBE value is slightly too small.

\begin{figure}[t]
\includegraphics[width=8.9cm]{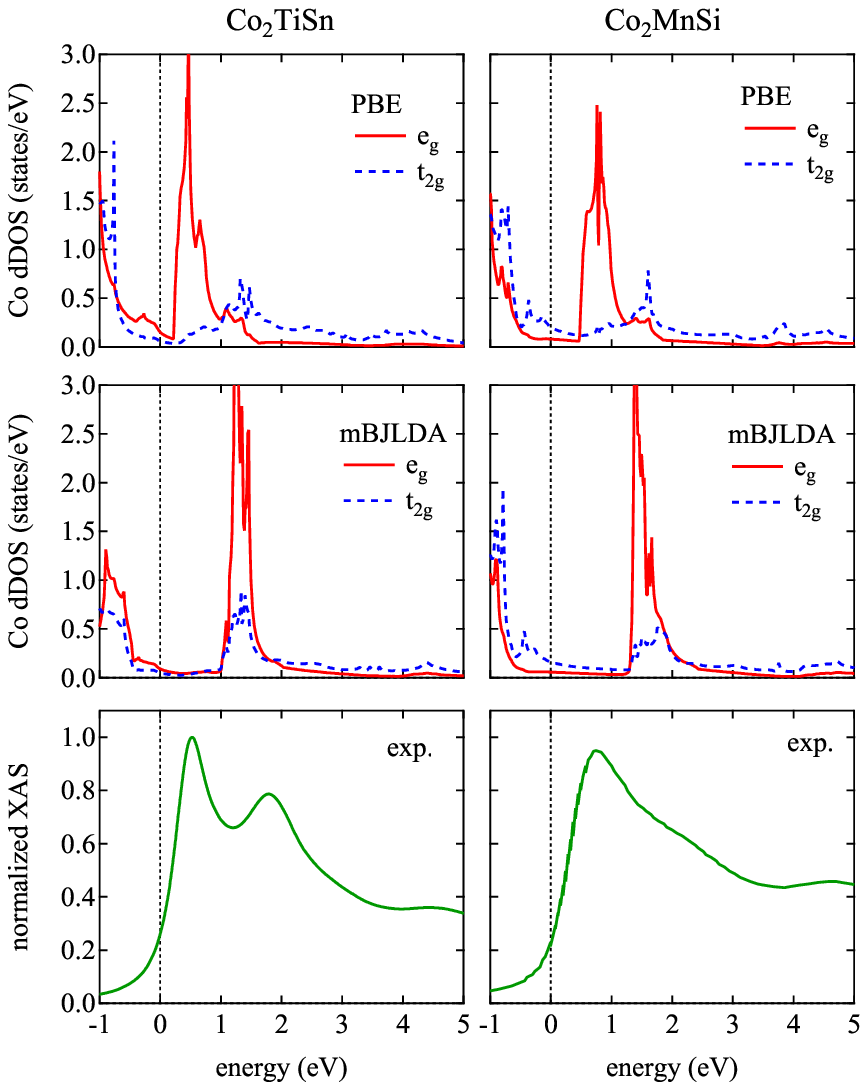}
\caption{\label{fig:CMS-CTS}Experimental Co L$_3$ absorption spectra of Co$_2$TiSn and Co$_2$MnSi (bottom row) and corresponding \textit{d}DOS with PBE and mBJ\-LDA approximations. Data taken from Refs. \onlinecite{Meinert11, Telling08}.}
\end{figure}

Also the spectral shapes of the Co L$_3$ x-ray absorption spectra are not well reproduced by mBJ\-LDA. Consider, e.g., Co$_2$TiSn (Fig. \ref{fig:CMS-CTS} left), which has a pronounced double-peak structure\cite{Meinert11} arising from a pure Co $e_g$ and a Co-Ti hybrid $t_{2g}$ state, which are separated by 0.9\,eV in the PBE calculation. Due to the excitation process, a screened core-hole is formed and pulls the localized $e_g$ states down by 0.3\,eV, so the total separation of the peaks is predicted to be 1.2\,eV, which is close to the experimental value of 1.3\,eV (see Fig. \ref{fig:CMS-CTS}). Note that the experimental spectra are broadened due to finite lifetime effects. This double-peak structure is clearly visible in the unoccupied PBE DOS of Co$_2$TiSn in Fig. \ref{fig:CMS-CTS}. In contrast, the mBJ\-LDA moves the $e_g$ peak up, such that it overlaps with the $t_{2g}$ peak, and no double-peak structure would be visible. A similar but less pronounced  structure is also present in the absorption spectra of Co$_2$MnSi (Fig. \ref{fig:CMS-CTS} right),\cite{Telling08} which is reproduced by PBE but is not present in the mBJ\-LDA calculation for the same reason as for Co$_2$TiSn. 

The gaps of Fe$_2$VAl and Fe$_2$TiSn predicted by mBJ\-LDA should easily be detectable with optical methods and by electrical transport. However, both compounds have been characterized as semimetals by experiments.\cite{Okamura00, Feng01, Dordevic02} Hence, the overestimation of the Heusler compound band gaps by mBJ\-LDA is not limited to magnetic cases, where the different spin densities may lead to an error in the determination of $c$, but is also found for paramagnetic materials.

While the band gap (and actually the entire band structure) is a property of excited states, the magnetic moments and hyperfine fields are ground state properties. The total moments were already discussed above and good agreement with experiment was shown for PBE and mBJ\-LDA in most cases. We now turn to the hyperfine fields $H_\mathrm{hf}$, which we calculate from the Fermi contact term, ignoring other contributions.\cite{Bluegel87} As mentioned above, the BJLDA potential ($c = 1$) provides densities of states that are very similar to those of PBE. Therefore we compare the hyperfine contact fields computed with PBE, mBJ\-LDA, and BJLDA for five Heusler compounds. The hyperfine fields and the corresponding site-resolved magnetic moments are given in Table \ref{tab:hyperfine}. The PBE functional provides reasonable hyperfine contact fields with a mean absolute error (mean error) of 64\,kOe ($+$44\,kOe) for the present test set. No clear trend towards an over- or underestimation of the hyperfine fields is visible. The PBE values are similar to those calculated by Picozzi \textit{et al.}\cite{Picozzi02} mBJ\-LDA and BJLDA perform less well, with mean absolute errors (mean errors) of 133\,kOe ($+$73\,kOe) and 135\,kOe ($+$101\,kOe), respectively, where the mean of the experimental values is $-$78\,kOe. Since mBJ\-LDA and BJLDA produce very different results, the hyperfine fields depend critically on the value of $c$. Again, no clear trend towards an over- or underestimation with respect to experiment or with respect to the other approximations is obvious. Although BJLDA and PBE predict similar magnetic moments and DOS, there are notable differences in the wavefunctions, particularly those of the \textit{s} states, which provide the dominant contribution to the transfered hyperfine contact field. PBE performs best in this respect.

\begin{table}[!t]
\caption{\label{tab:hyperfine}Hyperfine contact fields (in kOe) and site-resolved magnetic moments (in $\mu_\mathrm{B}$) of chosen Heusler compounds. See text for mean (absolute) errors.}
\begin{ruledtabular}
\begin{tabular}{ l r r l r r l r r l r }
& \multicolumn{ 2}{c}{PBE} &  & \multicolumn{ 2}{c}{mBJ\-LDA} &  & \multicolumn{ 2}{c}{BJLDA} &  & \multicolumn{1}{c}{exp.} \\ 

& $H_\mathrm{hf}$& $m$ &  & $H_\mathrm{hf}$ & $m$ &  & $H_\mathrm{hf}$ & $m$ &  & $H_\mathrm{hf}$\\ \hline
\multicolumn{11}{c}{Co$_2$MnSi} \\
Co & $-$108 & 1.07 &  & $-$139 & 1.18 &  & $-$45 & 1.12 &  & $-$145\footnotemark[1]\\ 
Mn & $-$219 & 2.85 &  & $-$154 & 3.01 &  & $-$51 & 2.91 &  & $-$336\footnotemark[1]\\ 
Si & 26 & $-$0.04 &  & 60 & $-$0.06 &  & 40 & $-$0.05 &  & --- \\ \hline

\multicolumn{11}{c}{Co$_2$MnSn} \\
Co & $-$139 & 1.00 &  & $-$200 & 1.17 &  & $-$83 & 1.06 &  & $-$156\footnotemark[1]\\ 
Mn & $-$224 & 3.02 &  & $-$178 & 3.19 &  & $-$44 & 3.06 &  & $-$344\footnotemark[1]\\ 
Sn & 89 & $-$0.03 &  & 288 & $-$0.04 &  & 133 & $-$0.03 &  & 105\footnotemark[1]\\ \hline

\multicolumn{11}{c}{Co$_2$TiSn} \\
Co & 52 & 1.05 &  & 14 & 1.27 &  & 106 & 1.09 &  & 21\footnotemark[2]\\ 
Ti & $-$82 & $-$0.03 &  & $-$137 & $-$0.19 &  & $-$85 & $-$0.04 &  & --- \\ 
Sn & 94 & 0.00 &  & 331 & 0.01 &  & 155 & 0.01 &  & 82\footnotemark[2]\\ \hline

\multicolumn{11}{c}{Mn$_2$VAl} \\
Mn & $-$76 & 1.45 &  & 7 & 1.59 &  & 26 & 1.46 &  & $-$99\footnotemark[3]\\ 
V & $-$100 & $-$0.84 &  & $-$314 & $-$1.01 &  & $-$249 & $-$0.85 &  & $-$64\footnotemark[3]\\ 
Al & $-$31 & $-$0.02 &  & $-$37 & $-$0.03 &  & $-$31 & $-$0.03 &  & $-$25\footnotemark[3]\\ \hline

\multicolumn{11}{c}{Cu$_2$MnSn} \\
Cu & $-$234 & 0.06 &  & $-$219 & 0.05 &  & $-$190 & 0.04 &  & $-$175\footnotemark[2]\\ 
Mn & $-$79 & 3.44 &  & $-$4 & 3.62 &  & 139 & 3.41 &  & --- \\ 
Sn & 488 & 0.00 &  & 535 & 0.00 &  & 539 & 0.00 &  & 196\footnotemark[2]\\
\end{tabular}
\end{ruledtabular}
\footnotetext[1]{Reference \onlinecite{Picozzi02}}
\footnotetext[2]{Reference \onlinecite{Endo76}}
\footnotetext[3]{Reference \onlinecite{Kawakami81}}
\end{table}

To give some more insight into the origin of the hyperfine fields, we break them down into core and valence contributions for Co$_2$MnSi, where the Co and Mn 3\textit{s} states are treated as core levels (Table \ref{tab:hff_corevalence}). Here one finds that both the core and valence contributions differ strongly among the three potentials. It has been shown that the core contribution scales directly proportional to the magnetic moment within the muffin-tin spheres.\cite{Picozzi02} Here, we find on average -130\,kOe/$\mu_\mathrm{B}$ for PBE, -68\,kOe/$\mu_\mathrm{B}$ for mBJ\-LDA, and -43\,kOe/$\mu_\mathrm{B}$ with BJLDA, with little difference between Co and Mn. However, Nov\'ak \textit{et al.} have shown by applying a self-interaction corrected potential to the core states, that the core contribution should actually be \textit{larger} in magnitude than that obtained with PBE.\cite{Novak03} Thus, the (m)BJLDA values are clearly worse than the PBE values.

\begin{table}[t]
\caption{\label{tab:hff_corevalence}Total contact hyperfine fields, valence, and core contributions for Co$_2$MnSi with PBE, mBJ\-LDA, and BJLDA.}
\begin{ruledtabular}
\begin{tabular}{lrrrlrrrlrrr}
 & \multicolumn{3}{c}{PBE}   &  & \multicolumn{3}{c}{mBJ\-LDA}  &  & \multicolumn{3}{c}{BJLDA}  \\ 
 &total & val. & core &  & total & val. & core &  & total & val. & core \\ \hline
Co 	&	$-$108 	& 32 		& $-$140 	&  & 	$-$139 	& $-$61 	& $-$78 	&  & 	$-$45 	& 	1 	&	$-$46 \\ 
Mn 	&	$-$219 	& 152 	& $-$371 	&  & 	$-$154 	& 55 		& $-$209 	&  & 	$-$51 	& 	79 	&	$-$130 \\ 
Si 	&	26 		& 28 		& $-$2 	&  & 	6 		& 62	 	& $-$2 	&  & 	40 		& 	41 	& 	$-$1 \\ 
\end{tabular}
\end{ruledtabular}
\end{table}

The mBJ\-LDA potential was originally designed as an empirical means to obtain the band gap directly as differences of Kohn-Sham eigenvalues.\cite{Kim10} Thus, the overall band structure is not intended to be improved over other semilocal approximations, such as the PBE functional. In other words, the mBJ\-LDA potential is neither designed to be a better approximation to the quasiparticle self-energy, nor is it designed to be a better approximation to the true Kohn-Sham potential,\cite{Kim10} which is reflected in the worse description of the hyperfine fields. As shown by Kresse \textit{et al.}, the mBJ\-LDA band dispersions can be even worse than those of PBE, for example in the case of optical absorption spectra of SiO$_2$.\cite{Kresse12} They also point out, that a local potential cannot simultaneously predict the gap and the band dispersions correctly. On the other hand, mBJ\-LDA seems to be beneficial for the local magnetic moments of strongly correlated materials\cite{Koller11} and for the oxygen \textit{K} edge energy-loss near-edge structure (ELNES) description of NiO.\cite{Hetaba12}

\section{Conclusions}

We have calculated magnetic moments and densities of states of 26 (inverse) Heusler compounds with the Tran-Blaha modified Becke-Johnson + LDA (mBJLDA) potential. In the half-metallic cases, the gaps are much wider than those obtained with the PBE functional. We have shown for some cases (for which sufficient experimental data are available) that mBJLDA does not improve the description of the half-metallic band gaps or band structures of these compounds with respect to the PBE functional. The original Becke-Johnson potential + LDA (BJLDA) predicts similar band structures as PBE, but with a slightly larger gap. Also the magnetic moments agree well with PBE. Such a description does seem to be more reasonable for materials with a metallic dielectric response, whereas highly correlated magnetic insulators have been shown to be overall better described by the mBJLDA potential.\cite{Koller11,Hetaba12} On the other hand, the hyperfine fields (as an important ground state property) predicted by mBJLDA and BJLDA are worse than those from PBE.

\acknowledgments
The author thanks G\"unter Reiss for helpful discussions and support, as well as the \textsc{elk} developers for their work. Financial support by the Deutsche Forschungsgemeinschaft ist gratefully acknowledged.

\end{document}